\documentstyle [graphicx,11pt] {article}

\makeatletter
\renewcommand{\section}{\@startsection{section}{1}{0in}
        {0.4\baselineskip}{0.1\baselineskip}{\Large\bf}}
\renewcommand{\subsection}{\@startsection{subsection}{2}{0in}
        {0.25\baselineskip}{-\baselineskip}{\large\bf}}
\renewcommand{\subsubsection}{\@startsection{subsubsection}{3}{0in}
        {0.1\baselineskip}{-\baselineskip}{\normalsize\bf}}

\makeatother
\begin{document}

%
%
%

\begin{center}
%
{\LARGE \bf 
Searches for astronomical neutrino sources and WIMPs with Super-Kamiokande}
\end{center}

\begin{center}
%
%
{\bf Super-Kamiokande Collaboration, submitted by A.~Okada$^{1}$}\\
{\it $^{1}$ Institute for Cosmic Ray Research, University of Tokyo, Kashiwa, Chiba 277-8582, Japan\\
}

\end{center}

\begin{center}
{\large \bf Abstract\\}
\end{center}
\vspace{-0.5ex}
  Searches for astronomical neutrino sources and weakly 
interactive massive particles (WIMPs) using the Super-Kamiokande 
detector have been performed. We select the neutrino-induced 
upward muon events for the first 4 years, which is already the 
world largest data sample, and look for statistically significant excesses 
compared to the atmospheric neutrino background. No excess has been 
found so far.
Limits of upward muon flux from various potential sources are obtained.
 Also limits of upward muon flux due to annihilations of WIMPs in 
the Earth core, the Sun and the Galactic center are obtained as 
a function of WIMP masses. 

\vspace{1ex}
\section{Introduction}
\label{intro.sec}

High energy neutrinos other than the atmospheric ones are 
expected to come from energetic astronomical objects, where 
protons and/or ionized nuclei are efficiently accelerated, and also from
special places where ultra-heavy particles are rich and can decay or 
annihilate. Examples of the former objects are pulsars, active galactic
nuclei, etc. where accelerated particles can interact with nearby 
gases or photons resulting in pion/kaon decays. Examples of the latter 
are the core of Earth, Sun,
Galactic center, etc. where relic WIMPs produced at the time of Big Bang 
are trapped via elastic scatterings and made annihilate with each 
other.  

 To find such high energy neutrinos being excess in specified directions, 
upward going muons produced in rock surrounding the Super-Kamiokande 
via charged current interactions of neutrinos are used. Assuming the same 
energy spectrum as that of atmospheric neutrinos, the average energies of 
neutrinos are ranging around 10GeV for muons stopping inside the detector 
and around 100GeV for those going out through the detector,
respectively, with muon track lengths $>$ 7m in the detector.

 Therefore the upward going muons keep directions of parent neutrinos
better than the muons produced inside the detector which have average 
energy of about 1GeV. As well known, higher energy neutrinos with their 
larger interaction cross section and longer path length of the resulting
 muons will compensate the disadvantage of their low flux, especially
 for a harder spectrum generally expected from an astronomical origin.

 The most likely candidate of WIMP is the neutralino assuming the
lightest supersymmetric particle of supersymmetric theories.
Current LEP data and cosmological constraints impose the allowed range of 
the neutralino mass between 50GeV and 600GeV~\cite{Ellis99}. 
This mass range matches 
the energies of the upward going muons ($>$ 1.6GeV). 

The analyses of the upward through-going muons and also the 
upward stopping muons at Super-Kamiokande for the study of the atmospheric
 neutrinos were already published~\cite{Upthr99, Upstp99}. 
Here we concentrate our interest on the astronomical  
neutrino sources and WIMPs obtain the corresponding flux limits.

\section{Detection of upward muons}
\label{Detect.sec}

The Super-Kamiokande detector is a 50~kton cylindrical water Cherenkov
calorimeter located at the Kamioka Observatory  $\sim$1000~m
underground in the Kamioka mine, Japan.  The detector
is divided by an optical barrier instrumented with photomultiplier tubes
(``PMT''s) into a cylindrical primary detector region (the Inner
Detector) with inward-facing 11,146 50cm PMTs and a
surrounding shell of water (the Outer Detector) equipped with
1,885 20cm PMTs which allows the tagging of entering and exiting
particles~\cite{Fukuda98b}. 

The cosmic ray muon rate at Super-K is 2.2~Hz. Because of this dominant
background, the downgoing muons can not be used for the present
analysis.  The trigger efficiency for a muon entering the detector with
momentum more than 200~MeV/$c$ is $\sim$100\% for all zenith angles.  
The nominal detector effective area
for upward coming muons with a track length \(>\) 7m in the Inner Detector is
$\sim$1200~m$^2$.

An event is classified as an upward going muon, if all the following
criteria have been satisfied.  
They are a) a muon event that has a clear entering
signature, b) particle track's zenith angle cosine being less or
equal to 0, and c) length of track inside the Inner detector being greater
than 7m.  

For each event, we obtain the arrival direction and time. After a visual
scan by two independent groups and a final handfit direction, 
1265 upward through-going and 
311 upward stopping muons have
been observed in 1138 live observing days that span between April of
1996 and Nov of 1999.
The handfits agree with each other within 1.5$^{\circ}$. 
More details of the data reduction procedures can be found in \cite{Fukuda98b}.

The above sample is contaminated by some downward going cosmic ray muons
close to the horizon due to the tracking angular resolution of the detector 
and multiple Coulomb scattering in the rock. The total numbers of such 
non-$\nu$ background events 
has been estimated to be 9.1 $\pm$ 0.8 for the upward through-going muons and 
21.4 $\pm$ 8.8 for the upward stopping muons, all contained in the
-0.1 $<$ cos${\theta} <$ 0  bin, where ${\theta}$ is the zenith angle. 
 The contamination from photoproduced upward going pions from
downward going cosmic ray muons is estimated to be $< 1 \% $~\cite{MACRO98}.

\begin{figure}[bh]
  \centerline{
    \begin{minipage}[t]{11cm}
       \includegraphics[width=11cm]{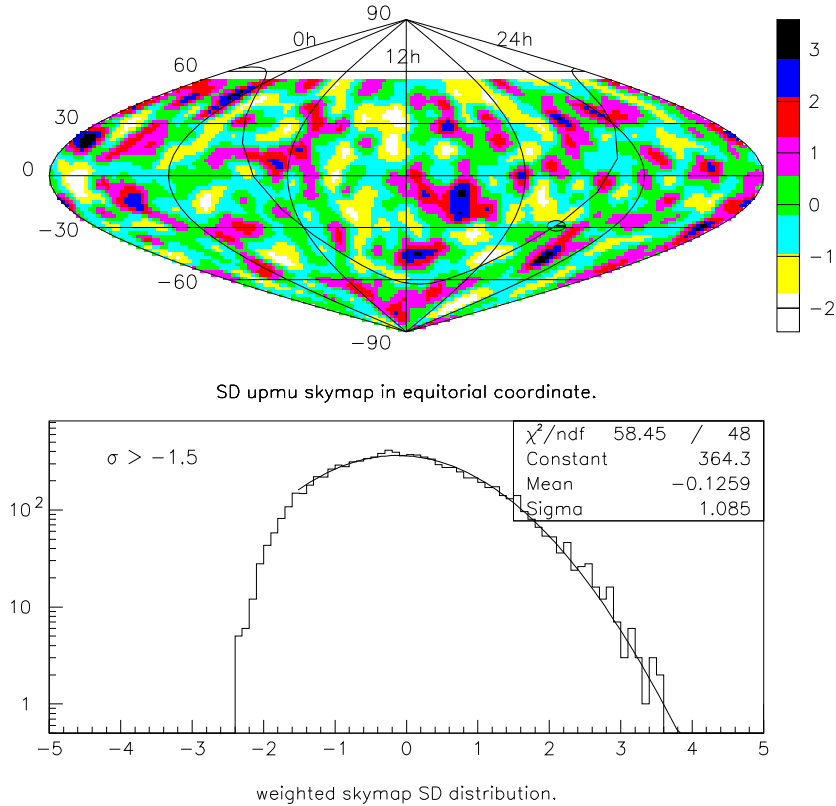}
    \end{minipage}
    }
  \caption{Sigma deviation sky map and SD distribution fitted to 
Gaussian for sigma $>$ -1.5 region.}
  \label{fig:SD}
\end{figure}
\clearpage

\section{Analysis for point source search}
\label{Analy.sec}

\subsection{Sigma deviation sky map}

To find out the direction(s) of the sky from which the upward muon events
show statistically significant excess, a sigma excess sky map is
generated.  The expected noise and its standard deviation for each
sky point is calculated using so called bootstrap method.  In this
method, the real data is used to generate faked sky maps by swapping
the event time and event direction randomly.  Because of an usage of
the real events, all the systematics due to upward muon angular
distribution, neutrino oscillation, live time unevenness in sidereal
day, etc. are all taken care of automatically.
However, if there is a strong astronomical neutrino source in the direction of
south pole, this method of using the real data could lead to an
underestimate of the source. We are under investigation of such effects
by a Monte Carlo simulation.  

To generate sigma deviation sky map, each upward muon is smeared with a
Gaussian distribution of 4 degree sigma. Here, 4 degree is the angular
deviation of a typical upward through-going muon from the direction of 
its parent neutrino. Then, weight contributions
of each event to entire sky points are calculated and summed up at
each point.  The resulting weighted sky map is compared with the
average noise weight and its standard deviation to calculate sigma
deviation of each sky point.  The result is shown in Figure~\ref{fig:SD}.

\begin{table}[h]
  \centerline{
    \begin{minipage}[t]{12cm}
       \includegraphics[width=12cm,height=12cm]{um752.epsi}
    \end{minipage}
    }
  \caption{List of upward muon flux limits obtained against various 
potential sources.}
  \label{tab:lim}
\end{table}

In the same figure, fit result of sigma deviation distribution to
the Gaussian above 1.5 sigma deficit is also shown.  As one can see
in this figure, a fit to Gaussian is reasonable in this area and
there is no data point above 3.6 sigma excess.  Therefore, there is
no sign of statistically significant excess in upward muon data in the sky.

\subsection{Flux limits from neutrino point sources}

Flux limits against various potential neutrino sources are obtained
by the following procedure.  First, the number of upward muons within 4
degree half angle cone from the sources is counted.  Then, 90 \%
Poisson confidence limit for that number is used to calculate flux
limit for the source.  Table~\ref{tab:lim} shows the flux limit results for the
selected potential neutrino sources, which are comparable with the latest 
data from the MACRO Collaboration~\cite{MACROastr}.  The expected noise in this
table is calculated using the same faked sky map method described
above, though no noise is subtracted to calculate the flux limit in
the table.

\begin{figure}[h]
  \centerline{
    \begin{minipage}[t]{12cm}
       \includegraphics[width=12cm]{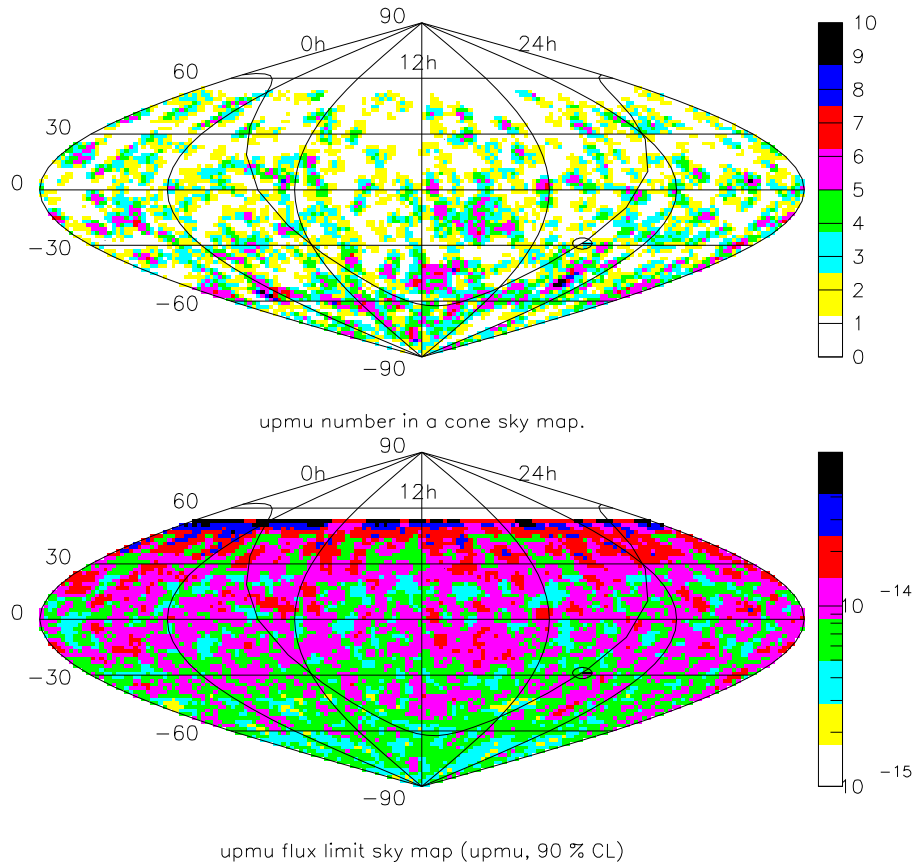}
    \end{minipage}
    }
  \caption{ Flux limit skymap calculated using 4 degree half angle
cone.  Top is number distribution of upward muons and bottom is
 converted to 90 \% CL flux limit in unit of cm$^{-2}$ s$^{-1}$.}

  \label{fig:skymap}
\end{figure}

\begin{figure}[h]
  \centerline{
    \begin{minipage}[t]{12cm}
       \includegraphics[width=12cm]{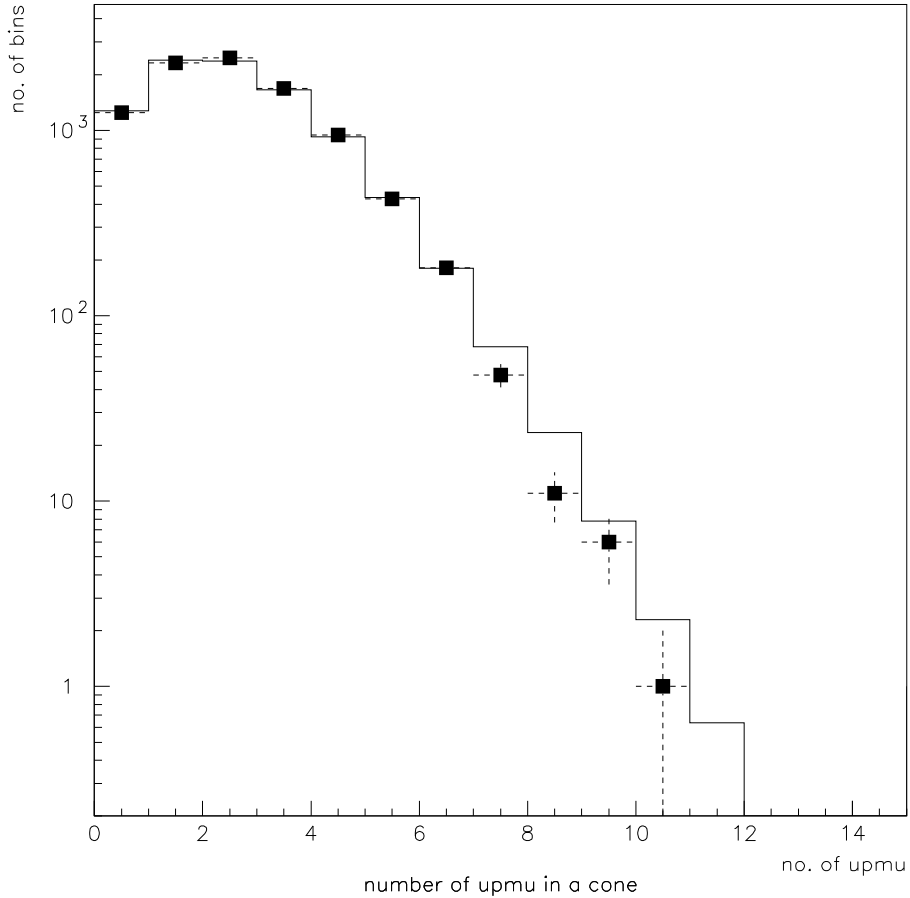}
    \end{minipage}
    }
  \caption{ Upward muon number distribution in 4 degree cone.  Solid histogram
shows noise distribution obtained with bootstrap method, and the black
circles show data.}

  \label{fig:numdis}
\end{figure}

The same method is employed to obtain flux limit for each of sky
points.  The resulting sky map is shown in Figure~\ref{fig:skymap}, along 
with a
figure showing the number of upward muons in the cone.  In the
Figure~\ref{fig:numdis}, the  event number distribution of upward muons in the 
cone is shown with
 that expected from noise.  This noise distribution is also obtained
 using the faked sky map method mentioned above.  The event number 
distribution of upward muons is consistent with noise, hence there is no 
sign of astronomical neutrino source in this method either.

\section{Analysis for WIMP search}
\label{Analy2.sec}

Recently the WIMP search at Super-Kamiokande was reported~\cite{Desai}. 
Here updated data are presented.

\subsection{Comparison of data with simulations of atmospheric neutrinos}

  The expected background for a WIMP search, which is due to 
interactions of atmospheric
$\nu$'s in the rock below the detector is evaluated with 40-year Monte Carlo
simulations. These simulations use
the Bartol atmospheric ${\nu}$ flux~\cite{Agarwal}, the GRV94 parton
distribution function~\cite{GRV94}, and energy loss mechanisms of muons in
rock  from~\cite{Lipari93}. There is a 20$\%$ uncertainty
in the prediction of absolute upward-going muon fluxes. 

\begin{figure}[hbt]
  \centerline{
    \begin{minipage}[t]{9cm}
       \includegraphics[width=9cm]{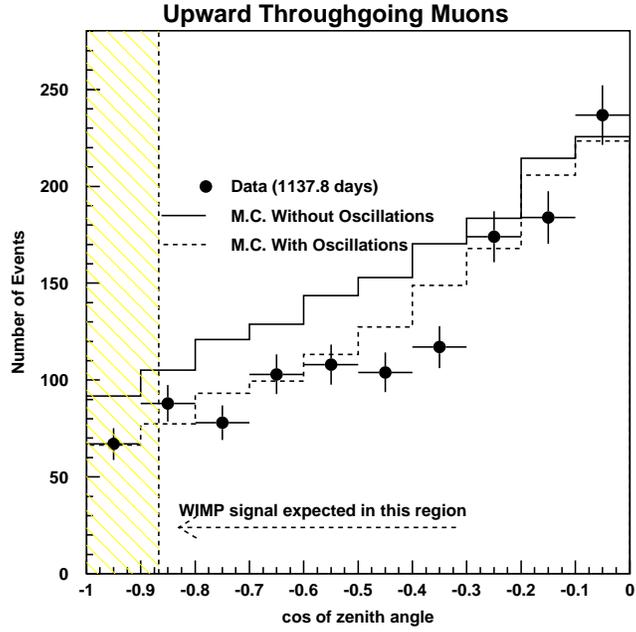}
    \end{minipage}
    }
  \caption[]{Distribution of upward through-going muons with respect to the
    Earth. The oscillation parameters used for the atmospheric neutrino
    Monte-Carlo are: $\sin^2 2{\theta} = 1.0$ and 
 $\Delta \mbox{m}^2 =2.5 \times 10^{-3} \mbox{eV}^{2}$.}
\label{fig-1} 
\end{figure}

   Analysis of the most recent  Super-Kamiokande data~\cite{SK00} of upward  
going muons and contained events is consistent 
with  ${\nu}_{\mu} \rightarrow {\nu}_{\tau}$ oscillations
with   $\sin^2 2{\theta} = 1$ and $\Delta \mbox{m}^2 =3.5 \times 10^{-3} 
\mbox{eV}^2$ being
the best fit values. Therefore  for evaluating our 
 background we  suppress the atmospheric muon neutrino flux due to  
oscillations from ${\nu}_{\mu}$ to ${\nu}_{\tau}$. 
The distribution of upward through-going muons with respect to the Earth
 is shown in Figure~\ref{fig-1}.  

\begin{figure}[th]
  \centerline{
    \begin{minipage}[t]{9cm}
       \includegraphics[width=9cm]{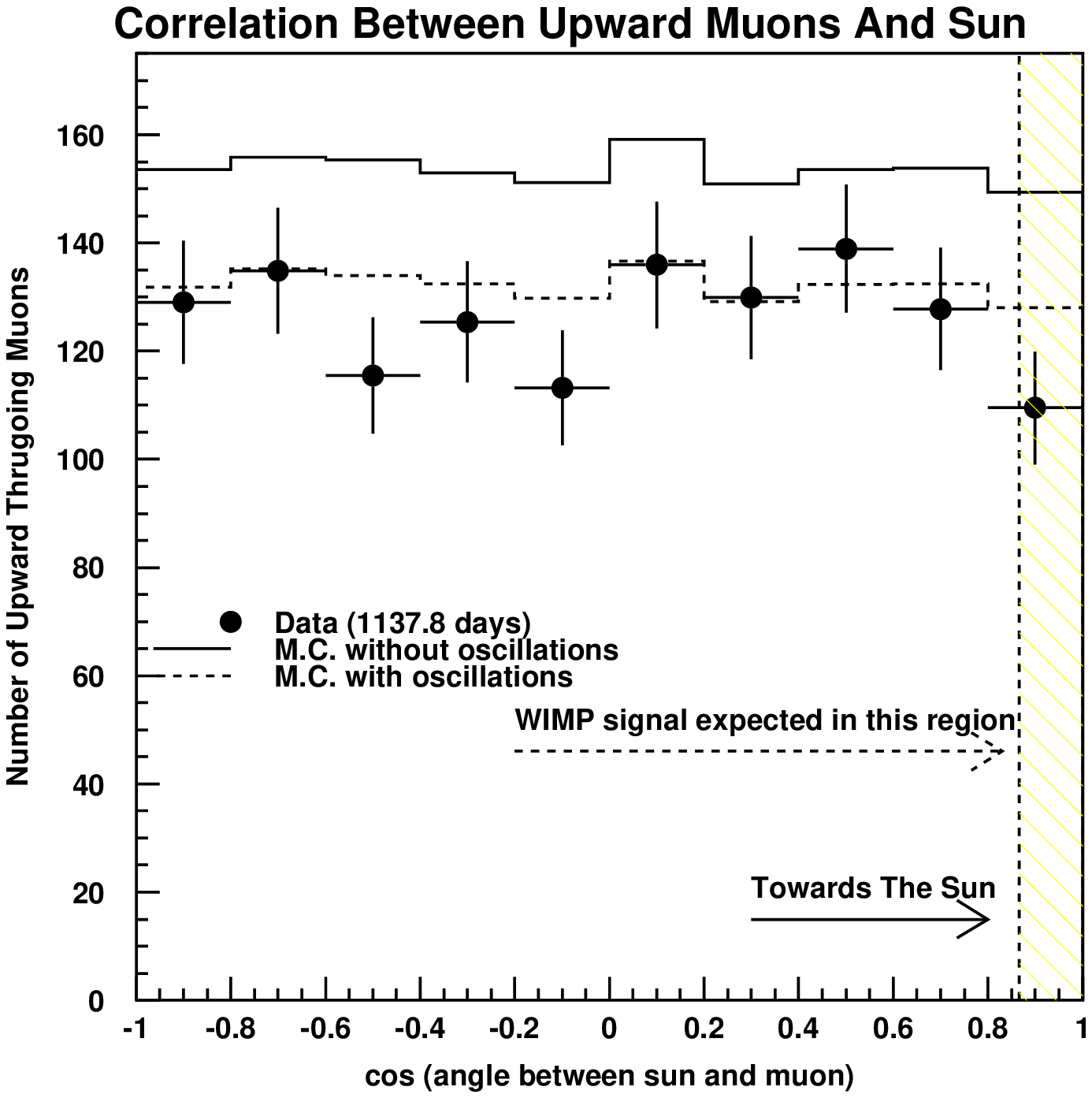}
    \end{minipage}
    }
\caption[]{Distribution of upward through-going muons with respect to the Sun
 The oscillation parameters used for the atmospheric neutrino
 Monte-Carlo are: $\sin^2 2{\theta}= 1.0$ and 
 $\Delta \mbox{m}^2 =2.5 \times 10^{-3} \mbox{eV}^{2}$.}
\label{fig-2} 
\end{figure}

\begin{figure}[tbh]
  \centerline{
    \begin{minipage}[t]{9cm}
       \includegraphics[width=9cm]{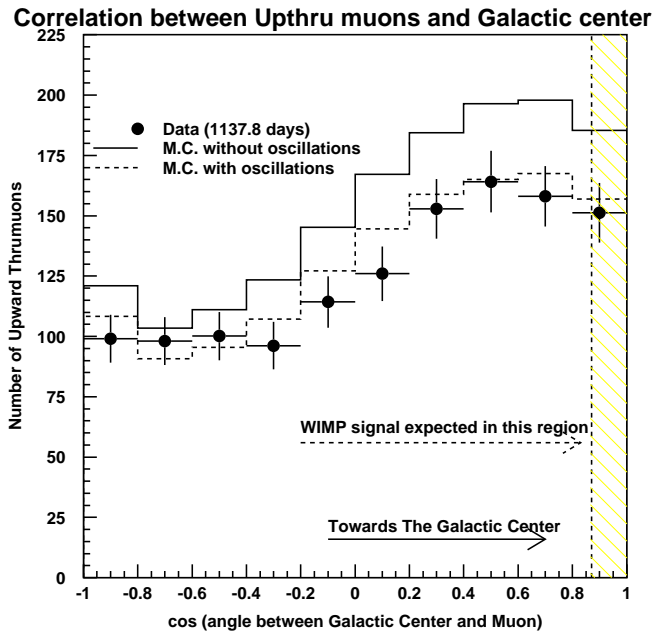}
    \end{minipage}
    }
\caption[]{Distribution of Upward Through-going Muons with 
respect to the Galactic Center. Coordinates of Galactic Center 
are: Right Ascension = 17h 42.4 m and Declination =$ -28^{\circ} 55' $. 
The oscillation parameters used for the atmospheric neutrino Monte-Carlo 
are: $\sin^2 2\theta = 1.0$ and  
$\Delta \mbox{m}^2 =2.5 \times 10^{-3}\mbox{eV}^{2}$.}
\label{fig-3} 
\end{figure}

    In order to compare the expected and observed distribution of 
upward through-going muon events with respect to
the Sun and Galactic Center, each Monte Carlo event was assigned random
times based on the arrival times of the observed upward through-going 
muon events. This procedure allows us to obtain  the angle between 
the upward muon
and any celestial object for each Monte Carlo event. 
The distribution of upward muons
with respect to the Sun and the Galactic Center is shown in
 Figures~\ref{fig-2} and~\ref{fig-3} respectively.
All the Monte Carlo events in Figures~\ref{fig-1},\ref{fig-2} and \ref{fig-3}
 are normalized by livetime.

\clearpage

\subsection {Flux limits due to WIMPs}

 We searched for statistically significant excess of muons 
in  cones with half angles ranging from 5 to 30 degrees.
This ensures that we catch about $90\%$ of the signal for a 
wide range of WIMP masses. Thus, searching in different cone angles
allows us to optimize the signal to noise ratio for various neutralino 
masses.

  No statistically significant excess was seen in any of the half 
angle cones. We calculate the flux limit of excess neutrino-induced muons  in 
each of the cones.

The flux limit is  obtained from the upper Poissonian limit (90$\%$
C.L.) given the number of measured events and expected background
\cite{PDB} due to atmospheric neutrinos, taking into account
oscillations.

 We also found that varying the oscillation
parameters and other normalization schemes hardly changes the flux limits 
close to the celestial objects. It is only in the  cone with half angle
  $30^{\circ}$ that the flux limits vary within 10$\%$ for different 
oscillation parameters. 

\begin{figure}[thb]
  \centerline{
    \begin{minipage}[t]{10cm}
       \includegraphics[width=10cm]{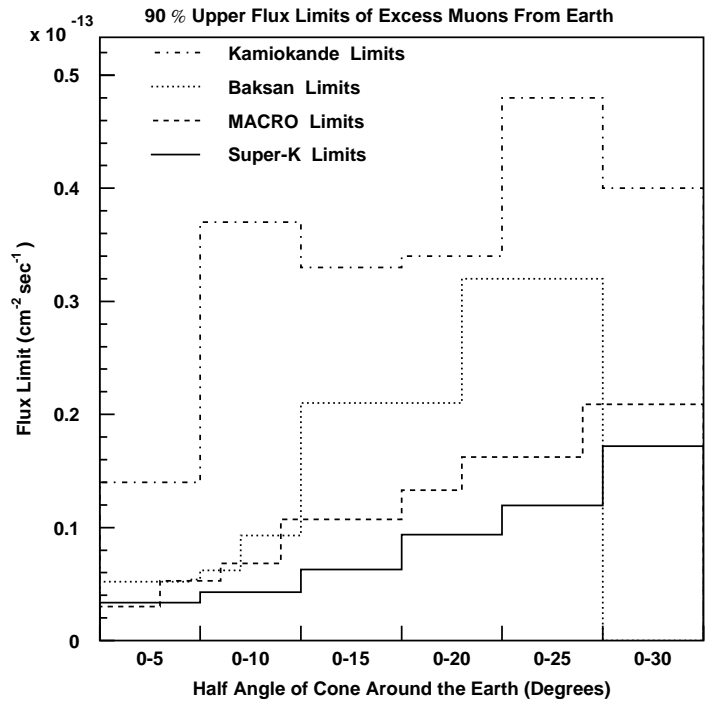}
    \end{minipage}
    }
\caption[]{Comparison of Super-Kamiokande excess neutrino-induced upward 
muon flux 
limits from the Earth with those from   other experiments.}
\label{fig-4} 
\end{figure}

\begin{figure}[thb]
  \centerline{
    \begin{minipage}[t]{10cm}
       \includegraphics[width=10cm]{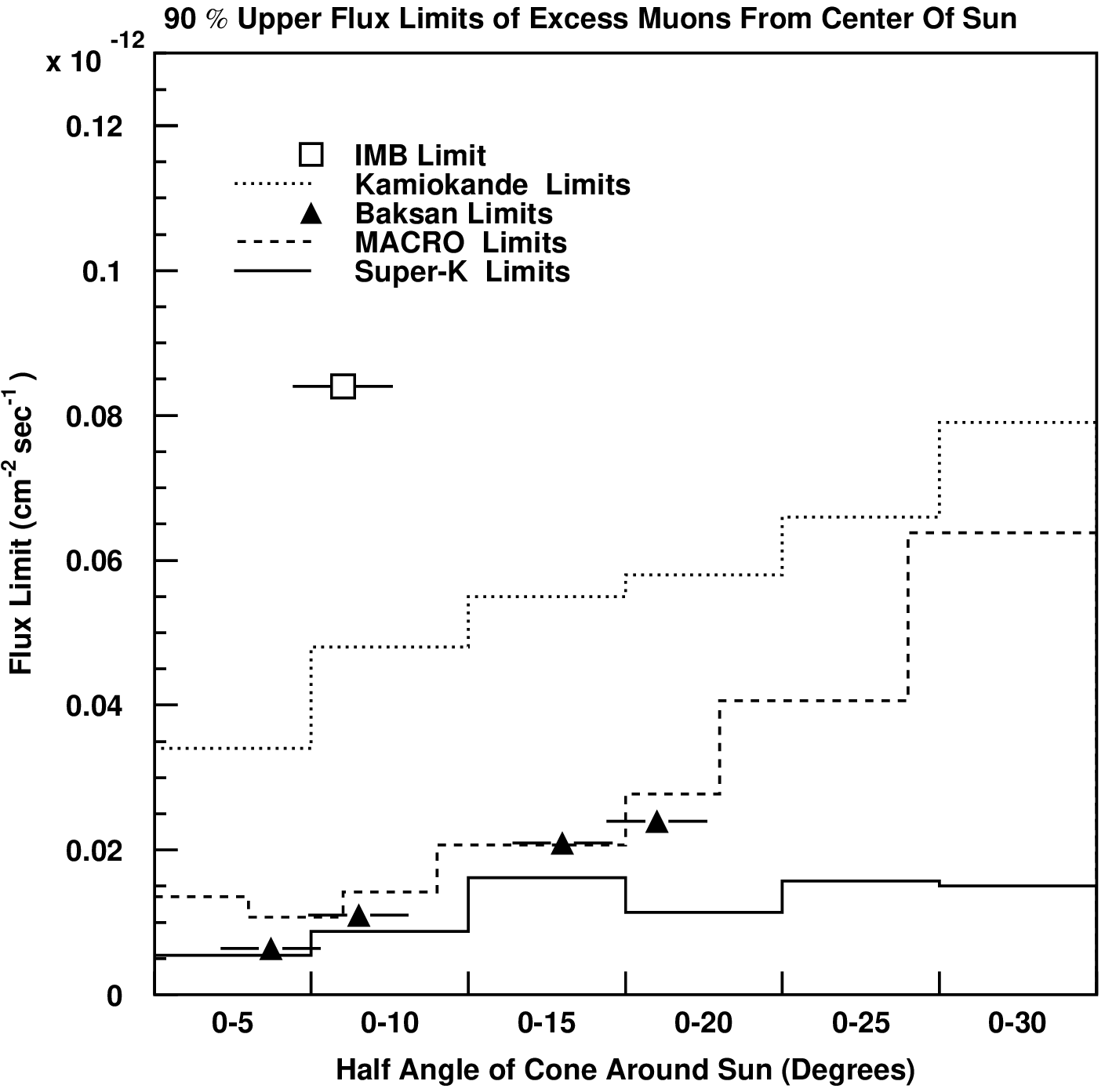}
    \end{minipage}
    }
\caption{Comparison of Super-Kamiokande excess neutrino-induced upward muon flux limits from  the Sun with those from  other experiments.}
\label{fig-5} 

  \centerline{
    \begin{minipage}[t]{10cm}
       \includegraphics[width=10cm]{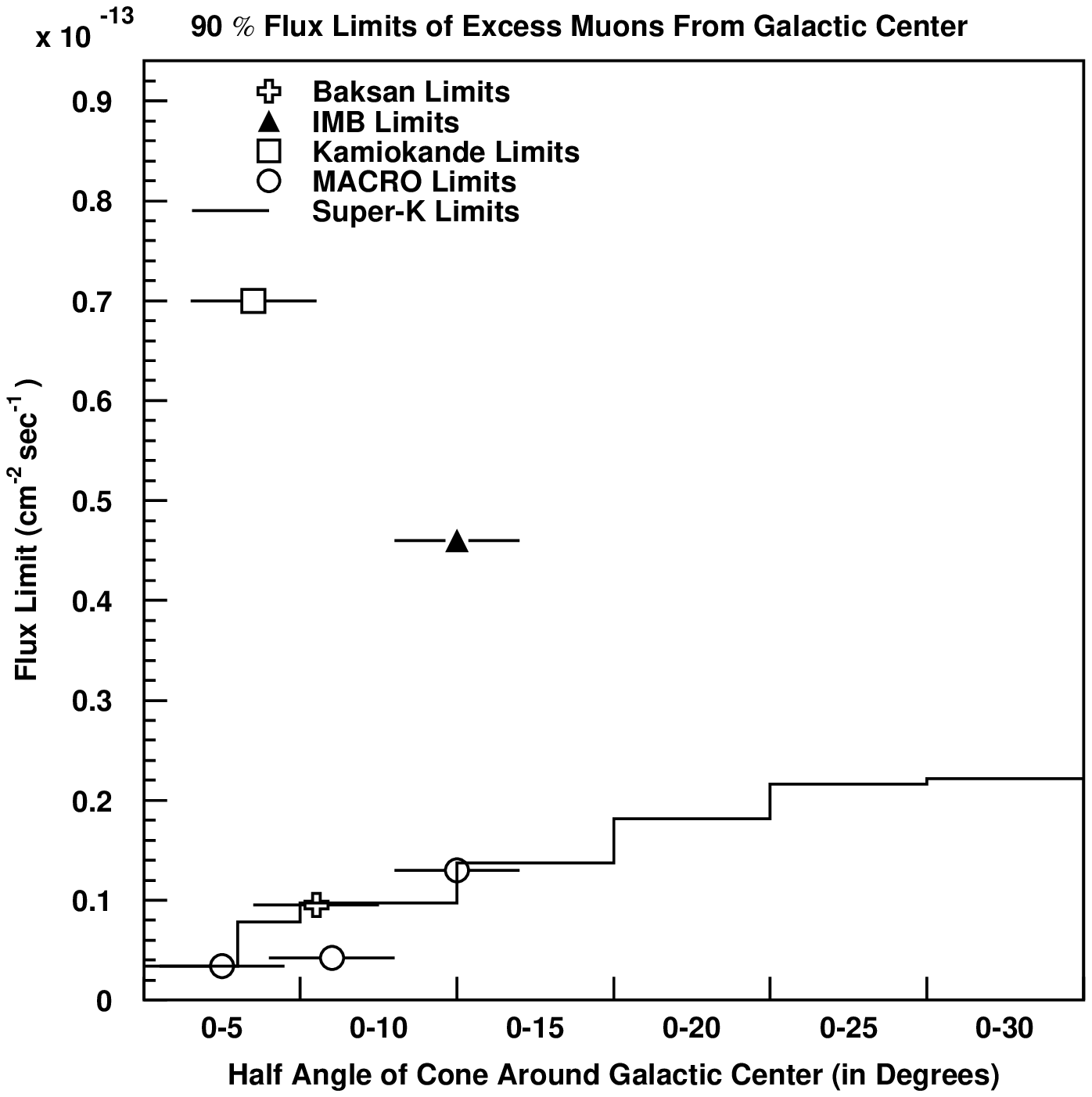}
    \end{minipage}
    }
\caption{Comparison of Super-Kamiokande excess neutrino-induced upward muon flux limits 
from the Galactic Center with those from other experiments.}
\label{fig-6} 
\end{figure}

  The comparison of Super-Kamiokande flux limits with previous estimates 
by other experiments is shown in  Figures ~\ref{fig-4},~\ref{fig-5}, and
~\ref{fig-6}  respectively. All the other experiments  have muon energy 
thresholds around 1 GeV. The WIMP flux limits for Earth and Sun by MACRO,
Kamiokande, Baksan,  and IMB are
in~\cite{MACROMi,Mori91,Baksan99,IMB86} 
respectively. The WIMP flux limits for the Galactic Center by MACRO,
Kamiokande, IMB, and Baksan are in ~\cite{MACRO00,IMB87,Kamiokande89,Baksan2}
 respectively.

\begin{figure}[thbp]
  \centerline{
    \begin{minipage}[t]{7cm}
       \includegraphics[width=7cm]{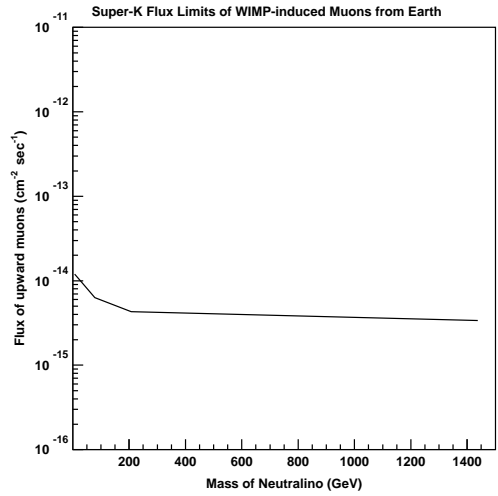}
    \end{minipage}
   }
  \centerline{
    \begin{minipage}[t]{7cm}
       \includegraphics[width=7cm]{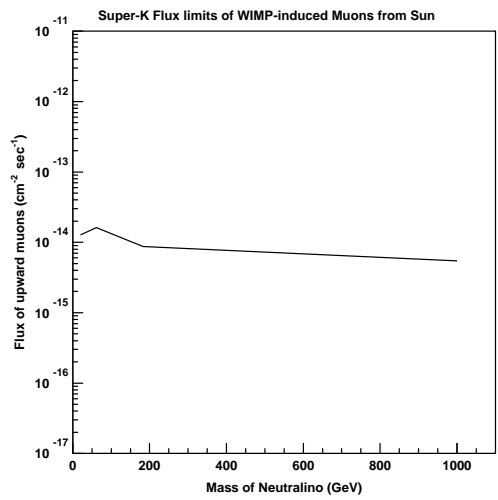}
    \end{minipage}
   }
  \centerline{
    \begin{minipage}[t]{7cm}
       \includegraphics[width=7cm]{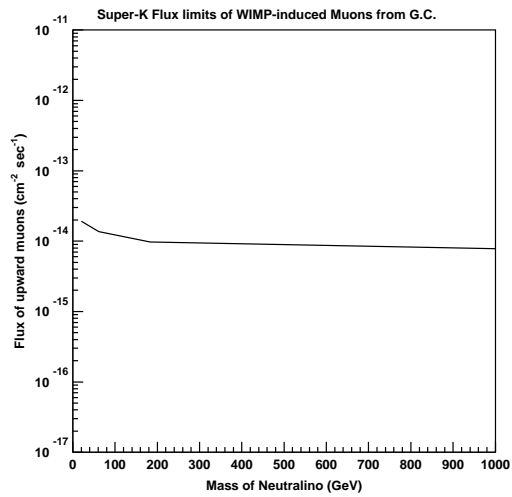}
    \end{minipage}
    }
\caption{ Super-Kamiokande WIMP-induced  upward-throughgoing muon flux limits from Earth (top), Sun (middle) and Galactic center (bottom) as a function  of  neutralino mass.}
\label{fig-7} 
\end{figure}

   Once WIMPs are captured in the Sun and the Earth they settle to the
core with an isothermal distribution equal to the core
temperature of the Sun or the Earth~\cite{Jungman96}. 
 The Kamiokande collaboration~\cite{Mori91} has calculated the angular
windows for Sun and Earth which contain 90$\%$ of signal for various
neutralino masses. These angular windows agree quite well with 
those obtained by the MACRO collaboration~\cite{MACRO99} for masses
down to 20 GeV. Using these windows, 90$\%$ confidence level flux limits can be
calculated as a function of neutralino mass using cones which collect
90 $\%$ of expected signal for any given mass. These flux limits as a 
function of neutralino mass are shown in Figure~\ref{fig-7} for Earth and
 Sun.

     Contrary to the Sun and Earth, the annihilation profile for the 
Galactic Center does not depend on the neutralino mass, because 
collisionless particle  dark matter 
does not come into thermal equilibrium near the Galactic 
Center~\cite{Gondolo}. However the apparent  size of the annihilation 
region is 
less than 0.05$^{\circ}$. Hence the Galactic Center can be considered a
point source for WIMP annihilation. Thus, similar to the Sun, one can
exploit the angular dependence of the charged current interaction  and multiple
coulomb scattering on neutrino energy to obtain the flux limits for 
the Galactic Center as a  function of mass. The results are shown also in 
Figure~\ref{fig-7} (bottom).

These flux limits can be compared with predictions from supersymmetry models.
Recently, the  DAMA/NaI experiment for direct detection of dark matter 
reported an annual modulation effect at 4$\sigma$ confidence
level which they claim is caused by WIMP scattering in their 
detector~\cite{DAMA00}. 
Their data is consistent with a relic neutralino forming
 a major dark matter component of the Galactic Halo~\cite{Bottino00}. 
The Super--Kamiokande flux limits could  be used
to rule out some regions of parameter space suggested by DAMA results,
These results shall be reported in a future publication. 

\vspace{1cm}

\clearpage

\clearpage

\section {Conclusions} 

  A search for astronomical neutrino point sources was done using neutrino 
induced 1265 upward
 through-going and 311 upward stopping muon events corresponding to 1138 days
 of livetime. 
  Also an indirect search for dark matter was done using the 1265 upward 
through-going muon events on the plausible assumption that  high energy 
neutrinos can be produced from  WIMP annihilation
 in the Sun, the Earth, and  the Galactic Center. 

We looked for an excess of
 upward muons
over atmospheric neutrino  background  in the direction close to the 
potential celestial objects.
No statistically significant excess was seen.

 Flux limits from almost all directions in the sky were obtained and the
definite numbers are given to various prominent astronomical objects. 
Also flux limits for various cone angles around the potential
WIMP sources were obtained   
and compared with previous estimates by other detectors.
 These flux limits can be calculated
as a function of the WIMP mass. 

{\bf Acknowledgments}

We gratefully acknowledge the cooperation of the
 Kamioka Mining and Smelting Company. The Super-Kamiokande experiment
 has been built and operated from funding by the Japanese Ministry of
 Education, Science, Sports and Culture, and the 
 United States Department  of Energy.

\end{document}